\documentclass[prl,superscriptaddress,twocolumn,amsfonts,amsmath,a4paper,showpacs]{revtex4}
\usepackage{graphicx}

\begin{document}

\title{A Novel Route to Calculate the Length Scale for the Glass Transition in Polymers }

%
%
\author{D. Cangialosi}
\email[Corresponding author:]{swxcacad@sw.ehu.es}
\affiliation{Donostia International Physics Center, Paseo Manuel de
Lardizabal 4, 20018 San Sebasti\'{a}n, Spain.}
\author{A. Alegr\'{i}a}
\affiliation{Dpto. de F\'{\i}sica de Materiales, Universidad del
Pa\'{\i}s Vasco (UPV/EHU), Apdo. 1072, 20080 San Sebasti\'{a}n,
Spain.} \affiliation{Unidad de F\'{\i}sica de Materiales, Centro
Mixto CSIC-UPV, Apdo. 1072, 20080 San Sebasti\'{a}n, Spain.}
\author{J. Colmenero}
\affiliation{Donostia International Physics Center, Paseo Manuel de Lardizabal 4,
20018 San Sebasti\'{a}n, Spain.}
\affiliation{Dpto. de F\'{\i}sica de Materiales, Universidad del Pa\'{\i}s Vasco (UPV/EHU),
Apdo. 1072, 20080 San Sebasti\'{a}n, Spain.}
\affiliation{Unidad de F\'{\i}sica de Materiales, Centro Mixto CSIC-UPV,
Apdo. 1072, 20080 San Sebasti\'{a}n, Spain.}

\begin{abstract}
The occurrence of glass transition is believed to be associated to
cooperative motion with a growing length scale with decreasing
temperature. We provide a novel route to calculate the size of
cooperatively rearranging regions CRR of glass-forming polymers
combining the Adam-Gibbs theory of the glass transition with the
self-concentration concept. To do so we explore the dynamics of
glass-forming polymers in different environments. The material
specific parameter $\alpha$ connecting the size of the CRR to the
configurational entropy is obtained in this way. Thereby, the size
of CRR can be precisely quantified in absolute values. This size
results to be in the range 1 $\div$ 3 nm at the glass transition
temperature depending on the glass-forming polymer.
\end{abstract}
\date{\today}
\maketitle

The nature of the glass transition is one of the most important unsolved problems
in condensed matter physics and research in this field has enormously intensified
in the last decades due to its strong fundamental as well as applicative implications.
Among the peculiar phenomena displayed by glass-forming liquids, the super-Arrhenius
temperature dependence of the viscosity and the structural correlation time is certainly
one of the most intriguing. In this framework, more than forty years ago Adam and Gibbs \cite{Adam}
theorized that such a pronounced temperature dependence of the structural correlation time
is due to a cooperative process involving several basic structural units forming
cooperatively rearranging regions (CRR), which size increases with decreasing temperature.
Since then a great deal of theoretical approaches \cite{Donth,Xia} as well as
simulation studies \cite{Donati} and very recently experimental studies
employing multipoint dynamical susceptibilities \cite{Berthier}
have been devoted in the search of good candidates for CRR as well as its size and temperature
dependence. All of these studies suggest that a growing correlation length with decreasing temperature
of the order of several nanometers exists.
According to the Adam-Gibbs (AG) theory of the glass transition,
the increase of the structural relaxation time with decreasing temperature,
accompanied by the growth of the cooperative length scale, is due to the decrease
of the number of configurations the glass-former can access, namely the configurational entropy ($S_{\rm c}$)
of the system. The connection between the structural relaxation time and $S_{\rm c}$ is expressed by \cite{Adam}:
\begin{equation}
\tau=\tau_0 \exp [C/(TS_c)],
\end{equation}
where $C$ is a glass-former specific temperature independent parameter and $\tau_0$
is the pre-exponential factor. The dynamics of a large number of glass forming systems
have been successfully described through the AG equation in both experiments for
low molecular weight glass formers \cite{Richert} and polymers \cite{Cangialosi},
as well as simulations \cite{Sastry}.
Apart from the relation between the relaxation time and the configurational entropy,
the AG theory provides a connection between the number of basic structural units
belonging to the CRR and the configurational entropy:  $N \approx S_c^{-1}$.
Several recent simulations studies, where the cooperative length scale and the configurational entropy
were obtained within the framework of the string-like motion and the potential energy landscape of the
glass-former respectively, successfully tested the validity of this correlation
\cite{Giovambattista}. Being the number of particles proportional to the volume of the CRR,
the characteristic length scale $\xi$ can be related to the configurational entropy by:
\begin{equation}
\xi/2=r_c=\alpha S_c^{-1/3},
\end{equation}
where $r_c$ is the equivalent radius of the CRR and $\alpha$ is a material specific parameter. Here,
the exponent $-1/3$ implies that
the fractal dimension for the growth of CRR equals 3. This value
means that CRR possess a compact shape growing in three directions. This hypothesis has been
rationalized within the framework of the random
first order transition
theory \cite{Stevenson}. Very recent molecular dynamics simulations
in a Lennard-Jones liquid, performed taking into account the motion of all particles
\cite{Appignanesi}, also suggest that the structural rearrangement involves the motion of compact regions.
Despite the ability to describe the temperature dependence of the structural relaxation and
predict an increasing length scale with decreasing temperature, the glass-former specific
parameter $\alpha$ is not reliably obtained from the AG theory as highlighted by several
authors \cite{Xia,Huth}. Hence the size of the CRR cannot be estimated a priori from the AG theory alone.
In this work we apply the AG theory to polymer blends and polymer-mixture systems in order
to provide a new route to precisely determine the parameter $\alpha$ and hence the size of CRR.
To do this, we incorporate the concept of self-concentration in the AG theory,
first proposed by Chung and Kornfield \cite{Chung} and later developed by Lodge and McLeish \cite{Lodge}
to describe the segmental dynamics of miscible polymer blends. This approach has been
recently exploited by us to describe the dynamics of miscible polymer blends and concentrated
polymer solutions and provided an accurate description of the component segmental dynamics
(the $\alpha$ relaxation) of these systems \cite{Cangialosi2}. The essential features of the concept of
self-concentration are explained in Fig. 1 and can be summarized as follows: when a volume
is centered on the basic structural unit of one of the polymers of a given blend,
the effective concentration ($\phi_{\rm eff}$) will be in general different from the macroscopic one.
Two extreme cases are possible: i) if this volume equals the volume of the
basic structural unit, then $\phi_{\rm eff}=1$; ii) if, on the other hand, this volume is large
enough, $\phi_{\rm eff}$ will be equal to the macroscopic concentration, $\phi_{\rm eff}=\phi$. The effective concentration
is related to the self-concentration ($\phi_{\rm s}$) and to the macroscopic concentration
through \cite{Lodge}:  $\phi_{\rm eff}=\phi_{\rm s}+(1-\phi_{\rm s})\phi$.
The self-concentration, namely the volume fraction occupied by the isolated chain containing
the unit of reference, can be related to the radius of the volume through simple geometric
considerations involving the Kuhn and the packing lengths \cite{Kant}. It is worth remarking that,
as the volume of CRR is expected to be neither large enough to make the concentration equal
to the macroscopic one nor equal to that of the monomer
($\phi<\phi_{\rm eff}<1$) \cite{Lodge,Cangialosi2,Kant},
the component dynamics will be intermediate between that of the pure polymer and the average structural
dynamics of the mixture. This means that the self-concentration concept constitutes
an extremely sensitive tool when exploring length scales of the order of those expected for CRR.

\begin{figure}
\includegraphics[width=1.05\linewidth]{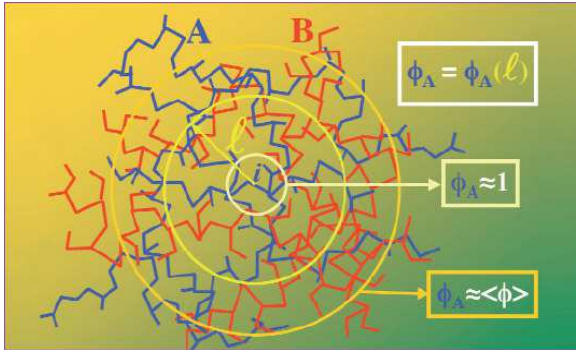}
\caption{(Color online) Schematic illustration of the
self-concentration concept.} \vspace{-2 mm} \label{figure1}
\end{figure}

In order to apply the AG theory to mixtures, $S_{\rm c}$, $C$, $\tau_{\rm 0}$ in Eqs. (1) and (2) have
to be evaluated in the cooperative volume. This means that these quantities are
function of the effective concentration of the polymer under examination.
In our previous works, we found that a linear combination of the parameters of
the pure components of the mixture through the effective concentration allows
a good description of dynamics data \cite{Cangialosi2}:
\begin{equation}
X=\phi_{\rm effA}X_{\rm A}+(1-\phi_{\rm effA})X_{\rm B},
\end{equation}
Where $X$ is either $S_{\rm c}$, $C$ or $\log \tau_{\rm 0}$. Eqs. (1), (3) and (4) can be fitted to
experimental data having $\alpha$ as the only unknown parameter. As this
is a glass-former specific parameter, the presence of the other
component in CRR can in principle affect the value of $\alpha$. This has
been found to be the case especially for polymer-solvent systems \cite{Cangialosi2}. We have
overcome this problem by studying the dynamics of polymer-polymer
and polymer-solvent mixtures at elevated concentration of the polymer
for which $\alpha$ is being calculated. This allows a straightforward extrapolation
of the $\alpha$ parameter to the pure polymer. Furthermore, it is noteworthy that
for concentrated polymer mixtures both concentrations fluctuations, which
might affect the dynamics, and deviations from ideality, which would induce
the failure of Eq. (3), are minimized. We have thus obtained values of $\alpha$ for
polystyrene (PS), poly(vinyl acetate) (PVAc) and used those already
obtained for poly(o-chlorostyrene) (PoClS) \cite{Cangialosi} and poly(vinyl methyl ether)
(PVME) ([22]). In such a way we have investigated four typical glass-forming
polymers with different rigidity. We have found that a single polymer specific
value of $\alpha$ is obtained when extrapolating to 100\% concentration independently
on the environment surrounding the polymer under consideration. The achievement
of a temperature independent   parameter results, via Eq. (3), in
an increasing length scale for structural relaxation with decreasing
temperature of the order of nanometers. In particular, the resulting
size of CRR was between 1 and 3 nm depending on the characteristic of
the polymer. For a given polymer in the accessible temperature
range the diameter of CRR only varies of about 20\%. PVAc, poly(2,6-dimethyl-1,4-phenylene oxide) (PPhO), ortho-terphenyl (OTP) and
toluene were purchased by Sigma-Aldrich. The molecular weight of the two
polymers were respectively $M_{\rm w}=83000g/mol$ ($M_{\rm w}/M_{\rm m}=4.2$)
and $M_{\rm n}=23000g/mol$ ($M_{\rm w}/M_{\rm n}=2.3$). PS was purchased from Polymer Source Inc. The molecular weight
was $M_{\rm n}=70400g/mol$ ($M_{\rm w}/M_{\rm n}=1.04$). Bis-phenol-C-dimethylether (BCDE) was
synthesized as describe elsewhere \cite{Meier}.
Highly concentrated PS and PVAc in different environments were investigated
by means of broadband dielectric spectroscopy (BDS). PVAc and PS dynamics were
investigated respectively in toluene and BCDE, and
OTP and PPhO. Dielectric
measurements were carried out using a high precision dielectric
analyzer (ALPHA, Novocontrol GmbH) and a Novocontrol Quatro cryosystem
for temperature control with a precision of ±0.1 K. Measurements were
performed over a wide frequency ($10^{-2}- 10^{6}Hz$) and temperature range in
isothermal steps starting for the highest temperature. As a general rule
the relevant relaxation time at any temperature was obtained from
dielectric relaxation spectra as the reciprocal of the angular
frequency at the maximum of the permittivity loss. Structural
relaxation data for toluene \cite{Doss}, BCDE \cite{Alvarez}, OTP \cite{Chang} and PPhO \cite{Robertson}
were taken from the literature.
Calorimetric measurements were carried out on pure PVAc, PS and BCDE by
means of the differential scanning calorimeter (DSC-Q1000) from
TA-Instruments. Measurements
were performed in temperature modulated mode with a average
heating rate of 0.1 K/min and amplitude of 0.3 K. Different
oscillation frequencies were investigated and the so-obtained
specific heats were extrapolated to zero frequency in order
to obtain quasi-static values of the specific heat itself. Specific
heats of toluene and OTP \cite{Yamamuro}, and PPhO \cite{Pak} were taken
from the literature. Since the configurational entropy cannot
be determined experimentally, the entropy of the liquid in
excess to the corresponding crystal was used exploiting the
proportionality between the two magnitudes \cite{Cangialosi}. This only scales
the value of $C$ in Eq. (1) and $\alpha$ in Eq. (3). A linear form
of the excess specific heat: $\Delta C_{\rm p}=C_{\rm p}^{melt} -C_{\rm p}^{crystal}$ , where $a$ and $b$
are material specific
constants, was employed to obtain the excess entropy \cite{Wunderlich}. This was
obtained integrating the relation: $\int_{T{_{\rm K}}}^{T} [\Delta C_{\rm p}(T)/T]dT$; where $T_{\rm K}$
is the Kauzmann temperature, namely the temperature where the entropy of the
liquid equals that of the crystal. Due to the lack of crystalline
specific heat data for glass-forming polymers, the latter quantity
was determined identifying it with the temperature where the
dielectric relaxation time of the $\alpha$ process tends to
diverge \cite{Richert}. Hence, the knowledge of the parameters $a$ and $b$ allows
fitting pure components dynamics data through the AG equation
(dashed line in Fig. 3 for pure PVAc) to obtain $C$, $\log \tau_0$ and $T_{\rm K}$.
The values of the parameters $a$ and $b$ together with $T_{\rm g}$, $T_{\rm K}$, $C$ and $\log \tau_{\rm 0}$
are summarized in table 1 for the pure polymers. Literature values of the
packing ($l_{\rm p}$) and Kuhn ($l_{\rm k}$) length, relating the size of CRR to the
self-concentration \cite{Kant}, are also included. In addition, previously
obtained BDS and calorimetric data for PVAc in poly(ethylene oxide) \cite{Tyagi},
for PoClS in PVME and low molecular weight PS \cite{Cangialosi2} and PVME in toluene,
PoClS and PS with various molecular weight \cite{Cangialosi2} were employed.
\begin{table}[tbh]
\begin{center}
\begin{tabular}{cccccccc}
&  &PVME  &PVAc &PS  &PoClS  \\
\hline
\\
& $T_{\rm g} (\rm K)$     & 249   & 304 & 373   & 402    \\
\vspace{0.2mm}
& $T_{\rm K} (\rm K)$     & 203.5   & 256 & 322   & 340    \\
\vspace{0.2mm}
& $\log \tau_{\rm 0}$     & -13.0   & -13.8 & -12.4   & -12.9    \\
\vspace{0.2mm}
& $C (\rm {Jmol^{-1}K^{-1}})$     & 21900   & 41450 & 26400   & 61800    \\
\vspace{0.2mm}
& $a (\rm {Jmol^{-1}K^{-1}})$     & 68.3   & 105.3 & 99.0   & 98.4    \\
\vspace{0.2mm}
& $b (\rm {Jmol^{-1}K^{-1}})$     & -0.16   & -0.21  & -0.17   & -0.17    \\
\vspace{0.2mm}
& $l_{\rm k} (\rm \AA)$     & 13   & 17  & 18   & 17.3    \\
\vspace{0.2mm}
& $l_{\rm p} (\rm \AA)$     & 2.7   & 3.7  & 3.9   & 3.9    \\
\vspace{0.2mm}
& $\alpha (\rm {J^{1/3}mol^{-1/3}K^{-1/3}})$     & $10.7 \pm 0.5$    & $17\pm 0.5$   & $19.5\pm 0.5$    & $24\pm 2$  \\
\\
\hline
\end{tabular}
\end{center}
\caption{Relevant parameters for all components of all
mixtures. Data for the $C$ parameter of the AG relation, the
parameters $a$ and $b$ of the excess specific heat and $\alpha$ are referred
to a per mole of monomer basis.} \label{tab:par}
\end{table}

In what follows, we will present the results obtained for mixtures
at high concentration of PVAc as a showcase and illustrate the methodology
followed to evaluate the parameter   for this polymer. Fig. 2 shows the
dielectric loss versus frequency for PVAc at various concentrations of
toluene and fixed temperature. The increase in toluene concentration
clearly provokes a speed-up of PVAc dynamics. Similar results were
obtained for PVAc in PEO and BCDE. The average relaxation time of PVAc
in mixtures with toluene as well as that of pure PVAc as a function of
the inverse temperature is displayed in Fig. 3. The continuous lines represent
the best fits of the proposed model to experimental data. The quality of the
fits is excellent. Analogously, the fits performed for the dynamics of PVAc in
mixture with PEO and BCDE resulted in a successful description of data. The
parameter $\alpha$ obtained from the best fits of our data is plotted in the inset
of Fig. 3 as a function of the average effective concentration. Strikingly
the extrapolation to pure PVAc leads to a single value of $\alpha$ independently
of the environment surrounding PVAc chains. This can be regarded as
the intrinsic $\alpha$ connecting PVAc configurational entropy with its size of
CRR. The same procedure was applied to evaluate the parameter $\alpha$ for PS
investigating its dynamics in OTP and PPhO and similarly to PVAc, a
single $\alpha$ was obtained. Furthermore, a single value of $\alpha$ was obtained
for PoClS and PVME studying their dynamics in various environments. The so-obtained
values of $\alpha$ are tabulated in table 1.
\begin{figure}
\includegraphics[width=1\linewidth]{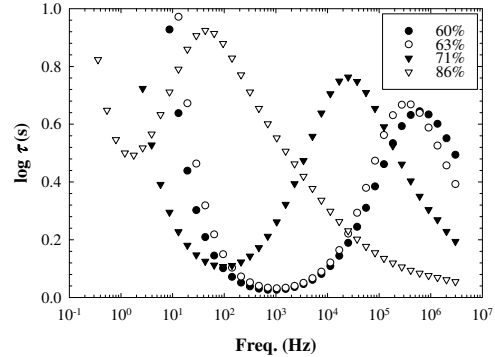}
\caption{Permittivity loss vs. logarithm of the frequency
for PVAc/toluene system at $308K$ and the following weight percentages
of PVAc: 60\% (filled circles), 63\% (empty circles), 71\% (filled triangles) and 86\% (empty triangles).}
\vspace{-2 mm}
\label{figure3}
\end{figure}
\begin{figure}
\includegraphics[width=1\linewidth]{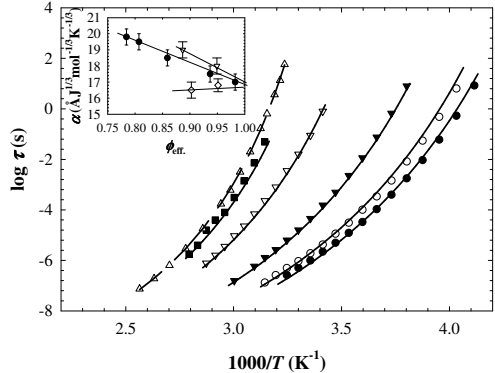}
\caption{logarithm of the relaxation time vs. temperature for PVAc
segmental dynamics in PVAc/toluene systems with the following PVAc
weight percentages: 60\% (filled circles), 63\% (empty circles), 71\%
(filled down-triangles), 86\% (empty down-triangles) and 95\% (filled
square). Pure PVAc segmental dynamics data are also shown
(empty up-triangles). The solid lines are the fits of the model to
PVAc segmental dynamics in toluene solutions and the dashed line is
the fit through the AG equation to pure PVAc dynamics data. The inset
of the figure represents the variation of the $\alpha$ parameter with the
average effective concentration of PVAc in toluene (filled
circles), BCDE (empty triangles) and PEO (empty circles). The solid
lines are a linear fit of experimental data, which, extrapolating to
pure PVAc, allow the evaluation of the $\alpha$ parameter for pure PVAc.}
\vspace{-2 mm}
\label{figure3}
\end{figure}

The availability of the polymer specific parameter $\alpha$ allows determining via equation (3) the
size of CRR and its temperature variation, starting from the knowledge of the excess entropy
determined from calorimetric measurements. The so-obtained diameter of CRR is
displayed in Fig. 4 as a function of temperature for the investigated
polymers. Though the size of CRR clearly depends on temperature, its variation
is rather smooth ($\approx20\%$ for the diameter). However, we note that this can translate
in a dramatic change of the number of particles involved in the cooperative
rearrangement. Furthermore, a clear relation with the rigidity of the chemical
structure can be observed, being the size of CRR larger for the most rigid
polymer, namely PoClS, and smaller for the most flexible, namely PVME. Remarkably
the addition of one bulky chlorine atom in the phenyl ring of PS results in a
substantial increase in the size of CRR for the resulting polymer,
namely PoClS. These results provide a positive indication of the
reliability of the approach followed. The dependence of the size of
CRR on the rigidity of the polymer suggests a positive correlation
between the size itself and the Kuhn segment, shown in table 1.
\begin{figure}
\includegraphics[width=1\linewidth]{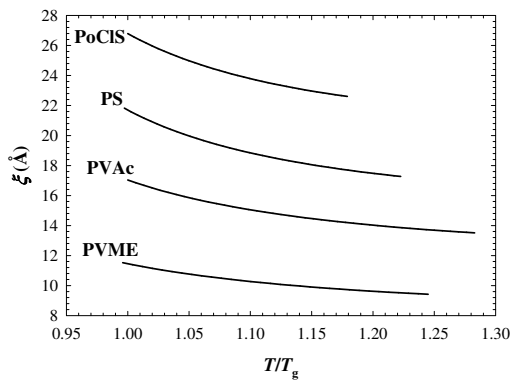}
\caption{Size of CRR vs. the temperature normalized at the $T_{\rm g}$ for all investigated polymers.}
\vspace{-2 mm}
\label{figure4}
\end{figure}
The assessment of the size of CRR allows performing a rough estimation of the number
of particles involved in the cooperative rearrangement. This can be done considering
the inter-chain distance for the investigated glass-forming polymers, which can be
obtained from the static structure factor $S(Q)$. Values for this distance can
be calculated from the available measurements of the static structure factor
of PS \cite{Furuya,Iradi}, PVAc \cite{Tyagi} and PVME \cite{Saelee}, being the resulting
values: $d_{\rm PS}=9 \rm \AA$, $d_{\rm PVAc}=8.4 \rm \AA$ and $d_{\rm PVME}=6.3 \rm \AA$. It is
noteworthy that the diameters of
CRR found at $T_{\rm g}$ for these polymers are almost quantitatively
twice the corresponding intermolecular distance, suggesting that
the cooperative rearrangement only involves the first shell around
a basic structural unit. This would imply that the number of
particles involved in each CRR at $T_{\rm g}$ is of the order of tens. The
possible connection between the size of CRR and the intermolecular
distance opens interesting scenarios in the rationalization of the
length scale involved in the structural relaxation of glass-formers, which
needs to be further explored.

Finally, it is worth comparing the sizes of CRR obtained here with the
AG approach with those previously estimated employing different
approaches. The size of CRR obtained for PVAc at $T_{\rm g}$, $\xi \approx 1.7 \rm nm$ agrees almost
quantitatively with the characteristic length scale obtained by
Berthier et al.\cite{Berthier} by means of multipoint dynamical
susceptibility ($\xi \approx 2 \rm nm$). The size of CRR obtained by us for PS is also
in qualitative agreement with that estimated by Hempel et al. \cite{Hempel} in
the framework of Donth approach \cite{Donth} ($\xi \approx 2.2 \rm nm$ vs. $\xi \approx 3 \rm nm$) and
in excellent agreement with that estimated by Sills et al. \cite{Sills} by means of nanoscale
sliding friction ($\xi \approx 2.1 \rm nm$). Our results are also compatible with the absence of finite
size effects in confinement down to 5 nm \cite{Kremer}.

To summarize, we have taken advantage of the concept of self-concentration in combination
with the AG theory of the glass transition to open a new route to obtain the size of
CRR and its temperature dependence for several glass-forming polymers. This new
approach proposed allows the unambiguous derivation of the polymer specific
parameter $\alpha$ connecting the size of CRR to the configurational entropy. Once $\alpha$ is
obtained for each polymer, the size of CRR in a wide temperature range
can be determined from the knowledge of the configurational entropy
of the pure polymer. The temperature variation of the diameter of CRR
for a given glass-forming polymer results to be of only about 20\% in
the accessible temperature range. Remarkably, the values of the diameter
of CRR at $T_{\rm g}$ for the investigated polymers closely correlate with both
the Kuhn length and the inter-chain distance.

The authors acknowledge the University of the Basque Country and Basque Country
Government (9/UPV 00206.215-13568/2001) and Spanish Minister of Education (MAT 2004-01617)
for their support. The support of the European Community within the SoftComp program is also acknowledged.

\end{document}